\documentclass[reprint,
superscriptaddress,
 amsmath,
 amssymb,
 aps,
]{revtex4-2}

\usepackage[utf8]{inputenc}

\usepackage{graphicx}
\usepackage{dcolumn}
\usepackage{bm}
\usepackage[colorlinks]{hyperref}
\hypersetup{
    linkcolor = magenta,   
    citecolor = blue,      
    urlcolor  = magenta    
}
\usepackage{color}
\usepackage{amsmath}

\begin{document}

\title{Instability and particle current control of a parametrically driven Bose-Einstein condensate in a ring-shaped lattice}

\author{L. Q. Lai}
\email{lqlai@njupt.edu.cn}
\affiliation{School of Science, Nanjing University of Posts and Telecommunications, Nanjing 210023, China}

\date{\today}

\begin{abstract}

We investigate the dynamics of a Bose-Einstein condensate in a one-dimensional ring-shaped lattice with the Peierls phase and site-dependent modulations, where the condensate is confined in a single deep trap and the interparticle interaction strength is modulated by a time-periodic driving field. The system has a finite spectrum, which limits the excitation regimes, and the Peierls phase typically induces imbalanced complex hopping amplitudes in each direction, leading to nonzero net particle currents along the lattice chain, which can hold nearly persistent even when the driving field is turned off after half of the period. The configuration provides a specific way for the coherent control of particle currents in many-body quantum systems with the help of an external driving field, and promotes the possible applications in future closed-loop atom circuits.

\end{abstract}

\maketitle

\section{Introduction}

A system of cold atoms, aside from leading to numerous interesting quantum phenomena, is of importance in quantum simulations of many-body systems and in the creation of novel configurations for atomtronic circuits. Ultracold atomic quantum gases in optical lattices have attracted particular attention both experimentally and theoretically \cite{morsch,eckardt1}, where all the parameters are highly controllable, and the state-of-the-art experimental techniques enable the intense explorations of relevant dynamics by explicitly manipulating the lattice potential, the periodicity, and the scattering length \cite{eckardt2,lignier,stoferle,fallani,goldman}.

The optical lattice has also been a promising testbed for the physical realization of fundamental Hamiltonians \cite{jaksch,greiner,bloch,aidelsburger,miyake,jotzu,das,wei}, where time-periodic driving is considered as a powerful tool for the coherent manipulations \cite{eckardt1}, making it feasible to enter regimes inaccessible in other condensed matter systems. In addition, different lattice geometries can be precisely produced by utilizing varying beams, among which researchers have implemented various ingenious methods, e.g., the coaxial electromagnetic coils and axially symmetric polarization elements, to generate the closed-loop trap and the ring-shaped optical lattice \cite{gupta,amico1,amico2,sakamoto,mao,zhao,henderson,ryu,dreismann,marti}, allowing for extensive studies on related physics in the typical geometry with periodic boundary conditions, such as persistent currents \cite{kolovsky,beattie,molina,oliinyk,cai,pradhan,xhani,guilleumas1}, phase transitions with modulated interparticle interaction strength \cite{kanamoto1,kanamoto2,qian,pu1,pu2,zhu,guilleumas2,mizuno}, the population dynamics \cite{jezek1,jezek2,jezek3,jezek4}, and rotational superflow and superfluid \cite{ramanathan,moulder,wright,amico3,guilleumas3,furutani,chergui}.

In the context of the ring-shaped configuration, exotic nonequilibrium dynamics of quantum atomic gases including superfluid-insulator transition and the dynamical suppression of matter-wave tunneling can be specifically revealed under the action of a time-periodic driving field \cite{eckardt2,lignier,stoferle}, especially when the hopping amplitudes are modified by the Peierls phase resulting from complex-valued tunneling matrix elements \cite{molina,shastry,bouzerar,roth,lienhard}. Coherently controlling the particle currents in such systems has recently aroused considerable interest, due to the experimental feasibility and the importance in the generation of future atomic devices, much of which has laid emphasis on homogeneous atomic gases or condensates subject to lattices based upon the Bose-Hubbard model. In the present work, we introduce a one-dimensional ring-shaped lattice in the presence of the Peierls phase, where a Bose-Einstein condensate is initially trapped in a single deep well and the scattering length is modulated periodically in time by an external driving field. We consider two cases of the site-dependent modulations, and analyze the regimes for parametrically exciting the system and the influences of typical Peierls phases on the decaying properties of the condensate. We also describe the structure of the particle jets and demonstrate the net particle currents when the driving field is turned off after half of the period, which can remain small but almost persistent.

The paper is organized as follows. In Sec.~\ref{model}, we introduce the one-dimensional ring-shaped lattice and employ the mean-field approach to describe the related equations of motion. In Sec.~\ref{weak}, we consider the weak-drive limit by taking the drive strength as small, and we figure out the regimes where the system can be parametrically excited. In Sec.~\ref{nonlinear}, we present the numerical analysis of the nonlinear behaviors, and further discuss the control of particle currents. A summary of the results and their potential applications are given in Sec.~\ref{summary}.

\section{Ring-shaped lattice and Mean-field equations}\label{model}

\begin{figure}[htbp]
\includegraphics[width=0.75\columnwidth]{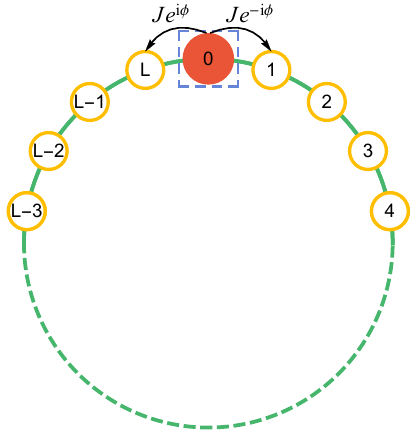}
\caption{Sketch of the ring-shaped lattice. The red solid circle $0$ represents the central site, which is confined in a trap symbolized by a blue dashed box. The orange empty circles labeled by integers $1,2,3,\ldots,L$ correspond to the sites outside of the trap. Hopping from site $j$ to $j+1$ has the matrix element $Je^{-i\phi}$, while hopping from $j+1$ to $j$ has the matrix element $Je^{i\phi}$.}
\label{ringlattice}
\end{figure}

As schematically illustrated in Fig.~\ref{ringlattice}, the system under consideration is a one-dimensional ring-shaped lattice, where a Bose-Einstein condensate is confined in the central site by a single deep well. The scattering length is modulated periodically in time by an external driving field, such that atoms with sufficient energies can be ejected from the trap, while traveling in opposite directions along the lattice chain. Under a weak time-periodic driving field, the density of atoms outside of the trap would be low when compared with that of the source condensate, and thus we neglect the interparticle interactions in that region, while only including the interactions between atoms which sit on site $j=0$ to keep the model simple. The system can be mathematically described by the Hamiltonian
\begin{eqnarray}
\hat{H} &=& V_{\rm{tr}}\hat{b}_{0}^{\dag}\hat{b}_{0}+\frac{1}{2}\left[U+g(t)\right]\hat{b}_{0}^{\dag}\hat{b}_{0}^{\dag}\hat{b}_{0}\hat{b}_{0}+\sum_{j=1}^{L}M_{j}\hat{b}_{j}^{\dag}\hat{b}_{j} \nonumber\\
&&-\sum_{j=0}^{L}\left(Je^{-i\phi}\hat{b}_{j+1}^{\dag}\hat{b}_{j}+Je^{i\phi}\hat{b}_{j}^{\dag}\hat{b}_{j+1}\right),
\end{eqnarray}
where $V_{\rm{tr}}$ is the trapping potential, and $\hat{b}^{\dag}_{j}$ $(\hat{b}_{j})$ is the bosonic creation (annihilation) operator at site $j$. The on-site pairwise interparticle interactions are characterized by a constant $U$ and a sinusoidal driving field $g(t)=g\sin(\omega t)$, with $g$ being the drive strength and $\omega$ being the drive frequency. $M_{j\geq 1}$ denotes the site-dependent modulation amplitude, and the hoppings between nearest-neighboring sites are quantified by complex amplitudes $Je^{\pm i\phi}$, with $\phi$ being the so-called Peierls phase. In experiments, the geometry can be possibly realized by optical lattices combined with a localized magnetic trap, where the interaction strength can be well modulated via the Feshbach resonance \cite{chin} and the Peierls phase may be simulated by rotating the ring-lattice trap \cite{das,he,fang,lacki,gorg,amico4}.

Basically, the source condensate in the single deep well and the resulting particle jets outside of the trap contain a macroscopic number of atoms; thus, at the mean-field level it is reasonable to replace the field operators with their expectation values $b_{j}=\langle\hat{b}_{j}\rangle$, where physically $|b_{j}|^{2}$ represents the number of particles on the $j$th site of the lattice chain. According to the expectation value of the Heisenberg equations of motion $i\hbar\partial_{t}\langle\hat{b}_{j}\rangle=\langle[\hat{b}_{j},\hat{H}]\rangle$, the corresponding discrete lattice Gross-Pittaevskii equations read
\begin{eqnarray}
i\hbar\dot{b}_{0}(t) &=& V_{\rm{tr}}b_{0}(t)+\left[U+g(t)\right]|b_{0}(t)|^{2}b_{0}(t) \nonumber \\
&&-Je^{i\phi}b_{1}(t)-Je^{-i\phi}b_{L}(t) \label{eom0}, \\
i\hbar\dot{b}_{j\geq 1}(t) &=& -Je^{-i\phi}b_{j-1}(t)-Je^{i\phi}b_{j+1}(t)+M_{j} b_{j}(t) \label{eomj}.
\end{eqnarray}
The periodic boundary condition is employed to identify $b_{L+1}$ with $b_{0}$, and the particle current flowing from site $j$ to $j+1$ can be defined as $I_{j}=Je^{i\phi}b_{j+1}^{*}b_{j}-Je^{-i\phi}b_{j}^{*}b_{j+1}$. When the external driving field is absent, i.e., $g=0$, the system is in the equilibrium state. Substituting the stationary ansatz $b_{0}(t)=\beta e^{-i\epsilon t}$, with $\beta$ being a constant, into Eq.~(\ref{eom0}) leads to ($\hbar=1$ hereafter)
\begin{eqnarray} \label{epsilon}
\epsilon=V_{\rm{tr}}+U|\beta|^{2}-Je^{i\phi}{\cal B}_{1}(\epsilon)-Je^{-i\phi}{\cal B}_{L}(\epsilon),
\end{eqnarray}
with, as derived in Appendix~\ref{gfunction},
\begin{eqnarray}
{\cal B}_{1}(\epsilon) &=& -Je^{-i\phi}{\cal G}_{11}(\epsilon)-Je^{i\phi}{\cal G}_{1L}(\epsilon), \\
{\cal B}_{L}(\epsilon) &=& -Je^{-i\phi}{\cal G}_{L1}(\epsilon)-Je^{i\phi}{\cal G}_{LL}(\epsilon),
\end{eqnarray}
where ${\cal G}_{mn}(\epsilon)$ is the frequency-domain Green's function. The nonlinear equation~(\ref{epsilon}) with respect to $\epsilon$ can readily be numerically solved, which indicates the appropriate parameter regimes for pumping atoms from the ground state to the corresponding excited state. In the following, we clarify the instabilities and the particle currents of the system under the action of the driving field.

 \section{Weak-drive limit} \label{weak}

When the drive strength $g$ is small, the driving field can be regarded as a perturbation. Thus, we employ the method of multiple scales to analyze the system, writing
\begin{eqnarray} \label{mom}
b_{0}(t)=e^{-i\epsilon t}\left[\beta(t)+g\chi^{*}(t)e^{i\omega t}+g\varphi(t)e^{-i\omega t}\right],
\end{eqnarray}
where $\beta(t)$, $\chi(t)$, and $\varphi(t)$ are slowly varying in time, and we assume that the slow-time variation of function $f(t)$ in the integral yields
\begin{eqnarray}
\int^{t}f(\tau)e^{-i\eta \tau}{\cal G}(t-\tau)d\tau \simeq f(t)e^{-i\eta t}{\cal G}(\eta).
\end{eqnarray}
Substituting Eq.~(\ref{mom}) into Eq.~(\ref{eom0}) and collecting the terms that are proportional to $e^{\pm i\omega t}$ result in
\begin{eqnarray} \label{matrixe}
\left(
\begin{array}{cc}
  \epsilon-\omega-\Lambda_{-}  & -U\beta^{*2}  \\
  -U\beta^{2}   &   \epsilon+\omega-\Lambda_{+}
\end{array}
\right)
\left(
\begin{array}{cc}
    \chi     \\
    \varphi
\end{array}
\right)
=\frac{i}{2}
\left(
\begin{array}{cc}
   |\beta|^{2}\beta^{*}   \\
   |\beta|^{2}\beta
\end{array}
\right)
\end{eqnarray}
with
\begin{eqnarray}
\Lambda_{\pm}&=&V_{\rm{tr}}+2U|\beta|^{2}+J^{2}{\cal G}_{11}(\epsilon\pm\omega)+J^{2}e^{2i\phi}{\cal G}_{1L}(\epsilon\pm\omega) \nonumber\\
&&+J^{2}e^{-2i\phi}{\cal G}_{L1}(\epsilon\pm\omega)+J^{2}{\cal G}_{LL}(\epsilon\pm\omega).
\end{eqnarray}

Along with the above equations, we also have
\begin{eqnarray}
i\partial_{t}\beta &=&2g^{2}|\beta|^{2}\left(\chi^{*}e^{i\omega t}+\varphi e^{-i\omega t}\right)\frac{e^{i\omega t}-e^{-i\omega t}}{2i} \nonumber \\
&&+g^{2}\beta^{2}\left(\chi e^{-i\omega t}+\varphi^{*}e^{i\omega t}\right)\frac{e^{i\omega t}-e^{-i\omega t}}{2i}.
\end{eqnarray}
Taking the ones that are winding, we obtain the time derivative of the total particle number as
\begin{eqnarray}
\partial_{t}|b_{0}(t)|^{2} &=& \beta\partial_{t}\beta^{*}+\beta^{*}\partial_{t}\beta \nonumber \\
&=&g^{2}|\beta|^{2}\left[(\beta^{*}\chi^{*}+\beta\chi)-(\beta^{*}\varphi+\beta\varphi^{*})\right]/2 \nonumber\\
&=&g^{2}|\beta|^{2}{\rm Re}(\beta\chi-\beta^{*}\varphi).
\end{eqnarray}
One can directly invert the matrix on the left of Eq.~(\ref{matrixe}) to find $\chi$ and $\varphi$. In the absence of the Peierls phase $\phi$, according to the frequency-domain Green's functions as derived in Appendix~\ref{gfunction}, for a uniform modulation amplitude $M_{j}\equiv M=-1$ the particle jets should be $0$ unless $-2J<\omega-|\epsilon|-M<2J$, while for a site-dependent amplitude $M_{j}=(-1)^{j}$ we have $|M|-2J\pm\sqrt{J^{2}+M^{2}}<\omega^{\prime}-|\epsilon^{\prime}|-|M|<\pm(|M|+2J)\mp\sqrt{J^{2}+M^{2}}$.

From a more fundamental point of view, the energy spectrum of a system without the site-dependent modulations is generally a continuum ranging from $-2J$ to $2J$. With typical modulation amplitudes, the spectrum has been renormalized, and it is required to obey the above restrictive conditions, i.e., to meet the energy conservation by applying appropriate driving fields, such that the particles can be pumped from the ground state to the corresponding excited state. As a result, a burst of atoms escapes from the trap into the continuum and moves along the lattice chain.

\section{Nonlinear dynamics} \label{nonlinear}

We are now in position to consider the nonlinear behaviors of the system by numerically solving the coupled equations (\ref{eom0}) and (\ref{eomj}) with the Runge-Kutta (RK4) method, where two cases of the modulation amplitudes $M_{j}\equiv M=-1$ and $M_{j}=(-1)^{j}$ are taken into account. The driving field is turned on at time $t=0$, and the energy is measured in units of $J$, such that the times and the frequencies are in units of $\hbar/J$ and $J/\hbar$, respectively. The numerics is performed by scaling $b_{j}\rightarrow b_{j}/\beta$, $U\rightarrow U|\beta|^{2}$, and $g\rightarrow g|\beta|^{2}$. Without loss of generality, we assume that $\beta=1$ and the trapping potential $V_{\rm{tr}}$ is negative, and in most of the numerics $J=1$.

\begin{figure}[htbp]
\includegraphics[width=\columnwidth]{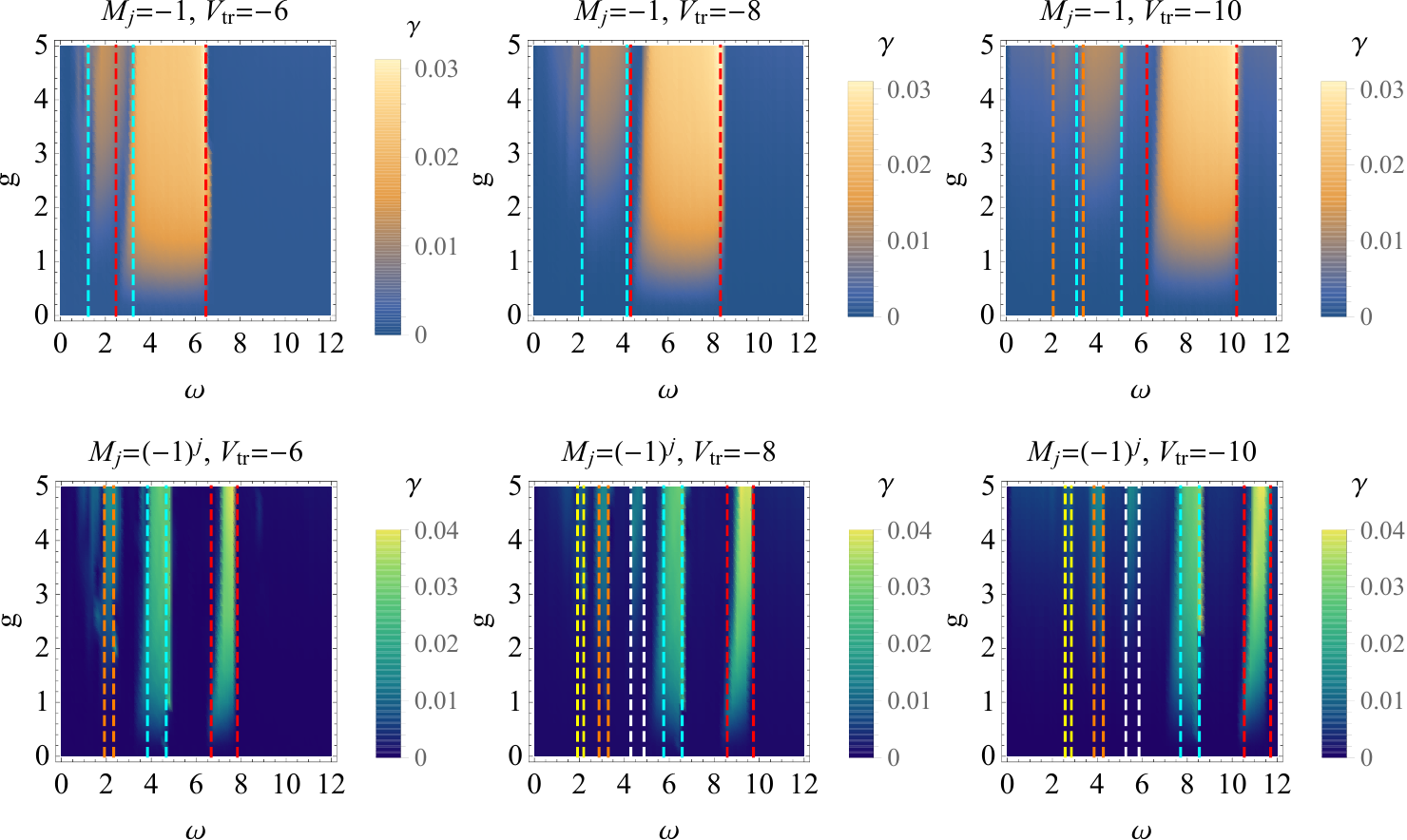}
\caption{(Color online) Average emission rate $\gamma$ of the particles from the condensate vs. drive frequency $\omega$ and drive strength $g$ under different trapping potentials $V_{\rm{tr}}=-6$, $-8$ and $-10$, with the Peierls phase $\phi=0$. The upper panel corresponds to the case of $M_{j}\equiv M=-1$, where the energies of the ground-state particles are $\epsilon_{6}=-5.47$, $\epsilon_{8}=-7.32$, and $\epsilon_{10}=-9.25$, respectively, while the lower panel is for $M_{j}=(-1)^{j}$, with the corresponding ground-state energies $\epsilon^{\prime}_{6}=-5.25$, $\epsilon^{\prime}_{8}=-7.17$, and $\epsilon^{\prime}_{10}=-9.12$. The dashed lines denote the characteristic scales that delineate the regimes for excitations, which is in accordance with Sec.~\ref{weak}. Here, we have taken the interaction strength $U=1$ and the number of lattice sites $L=100$.}
\label{decayratecom}
\end{figure}

We first figure out the regimes for parametrically exciting the system in the absence of the Peierls phase. The decay of the condensate would be rather nonexponential, especially when the on-site interaction strength $U$ and the drive strength $g$ are large. Nevertheless, one can fit the particle number in the central site to an exponential, i.e., $|b_{0}(t)|^{2}=Be^{-\gamma t}$, with $B$ being a constant and $\gamma$ representing the average emission rate of the particles, to quantify the emission process. Figure~\ref{decayratecom} clarifies how one could find expected excitations for different depths of the trapping potential $V_{\rm{tr}}$. The condensate can be quite stable for generic drive frequencies and drive strengths, unless $|\epsilon|-3<\omega<|\epsilon|+1$ for the case of $M_{j}\equiv M=-1$, and $|\epsilon^{\prime}|+\sqrt{2}<\omega^{\prime}<|\epsilon^{\prime}|-\sqrt{2}+4$ and $|\epsilon^{\prime}|-\sqrt{2}<\omega^{\prime}<|\epsilon^{\prime}|+\sqrt{2}-2$ for that of $M_{j}=(-1)^{j}$, which agree well with Sec.~\ref{weak}. Since the driving field also provides energy in multiples, at finite drive strengths there are further bands $\omega/n$ $(\omega^{\prime}/n)$ corresponding to $n$th order excitations, where they overlap for relatively shallow wells and separate for deeper ones. Though we specifically present the calculations at $U=1$, similar results can be found for different finite $U$. In what follows, we pay attention to the scenarios with moderately deep trapping potential and appropriate drive frequencies.

\begin{figure}[htbp]
\includegraphics[width=0.8\columnwidth]{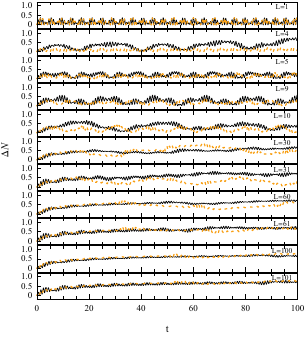}
\caption{(Color online) Time dependence of the number of ejected particles for different lattice sites $L$, with the Peierls phase $\phi=0$. The black solid lines and the orange dotted lines correspond to the cases of $M_{j}=(-1)^{j}$ and $M_{j}\equiv M=-1$, respectively. The trapping potential is $V_{\rm{tr}}=-8$, and the interaction strength is $U=1$. The driving field is applied with drive strength $g=2$ and drive frequency $\omega=6$.}
\label{jet9comtot}
\end{figure}

As a convenient model, one can definitely consider different situations with the number of lattice sites $L$ varying from small to large, and here we focus on the evolution of the particle number in the central site, where $\Delta N=|b_{0}(t=0)|^{2}-|b_{0}(t=t_{\rm{e}})|^{2}$ denotes the variation within a specific time $t_{\rm{e}}$. For the case of $M_{j}=(-1)^{j}$, one can see from Fig.~\ref{jet9comtot} the distinct behaviors with finite lattice sites $L$ and $L+1$: For $L=5$ and $L=9$ the number of ejected particles undergoes rapid oscillations in the whole driving process with a moderate maximum, while for the comparative $L=4$ and $L=10$ the maximums become somewhat larger, and they bear longer intervals before decreasing to almost $0$. With the increase of $L$, the deviations between odd and even lattice sites are greatly reduced. As for the case of $M_{j}\equiv M=-1$, the odd and even effects do not exist and the decaying properties simply depends on the number of the lattice sites. When the lattice sites are plentiful enough (e.g., $L=100$ and $L=101$), the behaviors can be quite similar to that of $M_{j}=(-1)^{j}$. Note that the results of the two cases can be nearly indistinguishable for $L=1$.

In general, when an appropriate driving field is applied, the ground-state particles in the condensate can be pumped to the corresponding excited state, and then be ejected from the trap to propagate along the lattice chain in both directions. If the number of lattice sites $L$ is small, the outgoing particles travel back to the central site instantly due to the ring-shaped geometry, and they collide with the source condensate before the reemission, such that $\Delta N$ oscillates rapidly with a moderate maximum. For the case with site-dependent modulation amplitudes $M_{j}=(-1)^{j}$, the system can be sensitive to the odd $j$ the even $j$ for $M_{2n}=1$ and $M_{2n-1}=-1$, and hence the decaying behaviors of $L$ and $L+1$ lattice sites deviate. The odd-even asymmetry might be further clarified in terms of the discrete parity symmetry. For example, when $L=5$ the correspondence would take place to the sites $1\leftrightarrow5$ and $2\leftrightarrow4$, where the counter-propagating excitations are symmetric, and site $3$ can be regarded as a ``mirror site" to the central site $0$ based on the left-right mirror in Fig.~\ref{ringlattice}. If the number of the lattice sites is either $L=4$ or $L=6$, however, the mirror site does not exist, and hence the symmetry is broken. Such symmetry breaking also leads to the progressive directed transport of interacting particles in ac-driven damped systems when the sources are explicitly distributed in the bulk \cite{malomed1,malomed2}. In contrast, the left-right excitations for the case of uniform amplitude $M_{j}\equiv M=-1$ remain symmetric regardless of the parity.

Moreover, when $L$ is larger it takes more time for the propagation, during which the emission of particles from the central site is uninterrupted, leading to a growing $\Delta N$ with a larger maximum. After a sufficient driving interval, the ejected particles return to the central site and $\Delta N$ also drops. With respect to the simplest case of $L=1$, there exist $M_{1}=-1$ in both cases, and hence they share analogous decaying properties. Note that $\Delta N$ would no longer precisely reach $0$, as there are always certain particles occupying the sites outside of the trap, especially when $L$ is large.

\begin{figure}[htbp]
\includegraphics[width=\columnwidth]{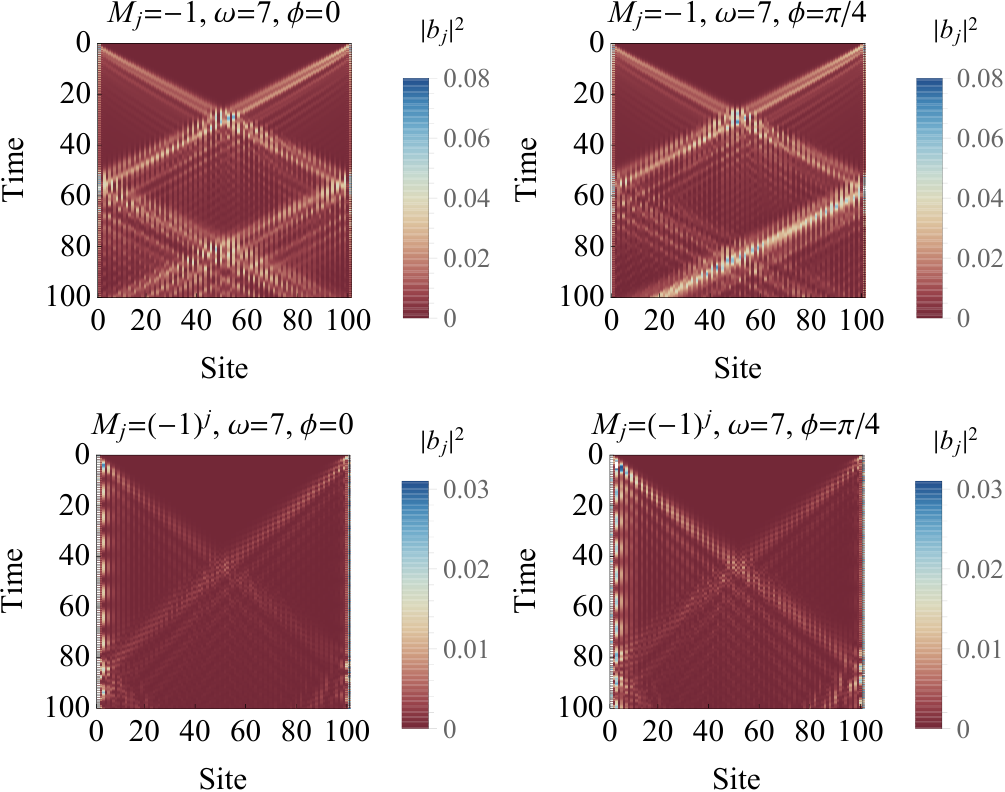}
\caption{(Color online) Time evolution of the number of particles on site $j$ for a typical drive frequency, with the Peierls phase being $\phi=0$ and $\phi=\pi/4$, respectively. The upper panel corresponds to the case of $M_{j}\equiv M=-1$, and the lower panel corresponds to that of $M_{j}=(-1)^{j}$. The trapping potential is $V_{\rm{tr}}=-8$, and the interaction strength is $U=1$. The drive strength here is $g=2$.}
\label{jetarraytot}
\end{figure}

The presence of a Peierls phase induces imbalanced complex hopping amplitudes $Je^{i\phi}$ and $Je^{-i\phi}$, which explicitly affects the emission process and renormalizes the boundaries of the excitation regimes. Basically, the exponential has generic equivalencies, e.g., $e^{\pm i\pi/4}=e^{\mp i7\pi/4}$, $e^{\pm i\pi/2}=e^{\mp i3\pi/2}$ and $e^{\pm i3\pi/4}=e^{\mp i5\pi/4}$, leading to very similar decaying behaviors of the condensate with related phases. Thus, one can routinely trace the particles on each site by demonstrating the number of particles $|b_{j}|^{2}$ as a function of time. Figure~\ref{jetarraytot} visualizes the structures of the particle jets with typical Peierls phases $\phi=0$ and $\phi=\pi/4$. Though the frequency $\omega=7$ appears to fall beyond the finite support of the spectrum of the case $M_{j}=(-1)^{j}$ based upon the regimes in Fig.~\ref{decayratecom}, and the resulting particle jets can be quite small, we take it here as an example for a better illustration of the phase effect. It is clear that when the Peierls phase is $\phi=0$, under the same driving conditions the jets for the case of $M_{j}\equiv M=-1$ can be much larger than those of $M_{j}=(-1)^{j}$, and the particles also travel faster. The counterpropagating particles along the lattice chain resemble a trace of ``back and forth" for both cases, where after reaching the intermediate sites they continue to move along the chain, and then collide with the source condensate before the re-emission. When the Peierls phase is tuned to $\phi=\pi/4$, the particle jet flowing through the $1$st site is correspondingly enhanced, while the one coming from the $100$th site is relatively weakened, which can be employed for the generation and control of the particle currents.

\begin{figure*}[htbp]
\includegraphics[width=\textwidth]{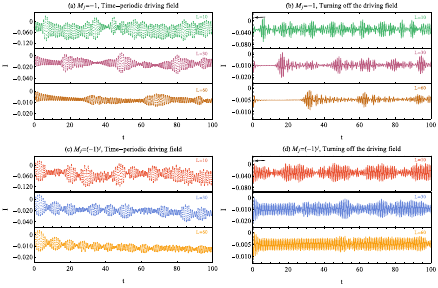}
\caption{(Color online) The net particle currents of lattice sites $L=10$, $30$, and $60$ for the cases of $M_{j}=-1$ and $M_{j}=(-1)^{j}$, respectively, with the time-periodic driving field being maintained [panels (a) and (c)] and the driving field being turned off at half of the period [panels (b) and (d)]. The black dashed lines in panels (b) and (d) denote the time $t_{\rm{h}}=\pi/\omega$ where we turn off the driving field. Here, we have taken $V_{\rm{tr}}=-8$, $g=2$, $\omega=6$, and $\phi=\pi/4$.}
\label{pctotcom}
\end{figure*}

In this scenario, we further try to turn off the external driving field at half of the period $t_{\rm{h}}=\pi/\omega$ with the Peierls phase $\phi=\pi/4$, and we illustrate the influences by comparing with the situation under an uninterrupted driving field, where the net particle currents induced by the Peierls phase are calculated from $I=\frac{i}{L}\sum_{j}I_{j}$. As can be plainly seen in Figs.~\ref{pctotcom}(a) and \ref{pctotcom}(c), when the time-periodic driving field is maintained, for both cases the net particle currents oscillate around typical ``negative" values for different lattice sites $L$ due to the Rabi oscillation between the ground state and the corresponding excited state \cite{mizuno}. The amplitudes of the resulting currents grow a bit with time, and basically the flows would be anti-clockwise according to the schematic of Fig.~\ref{ringlattice}, where one can simply reverse the direction by either shifting the Peierls phase from $\phi$ to $-\phi$, or making use of the generic equivalences of the exponential. When the driving field is turned off at $t_{\rm{h}}=\pi/\omega$, as shown in Figs.~\ref{pctotcom}(b) and \ref{pctotcom}(d), the net particle currents for the case of $M_{j}=(-1)^{j}$ appear to maintain the rapid oscillations throughout, while for $M_{j}\equiv M=-1$ there can be almost unchanged currents with fairly slow variations during certain intervals, and they both remain small but nearly persistent.

Under an uninterruptedly time-periodic driving field, the particle emission from the trap is actually continuous rather than transient, so there are always outcoming particles when the foregoing ones are returning, and the amplitudes of the net particle currents somewhat increase as the imbalances accumulate. However, it is not the case after the driving field is turned off. The system can remain in the excited state thereafter, but the particles are no longer continuously ejected from the trap, and the net particle currents are mostly generated by those occupying the sites outside, which explicitly limits the amplitudes at where the driving field is suspended. For a finite $L$, it takes a certain time for the propagation of particles along the lattice chain before they return to the central site, and hence there can be almost unchanged net particle currents with slow oscillations during the interval for the uniform case of $M_{j}\equiv M=-1$. After the collisions with the source condensate and the re-emission, the net particle currents oscillate rapidly again. As for the case of $M_{j}=(-1)^{j}$, due to the different modulations on each site, the net particle currents correspondingly vary during the propagation, exhibiting the rapid oscillations with the finite amplitudes throughout.

\section{Summary and Outlook}\label{summary}

We have introduced a one-dimensional ring-shaped lattice with the Peierls phase to explore the instabilities and the particle current control. A Bose-Einstein condensate is confined in the central site by a single deep well, where the interparticle interaction strength is modulated in time by a sinusoidally oscillating driving field. We consider two cases of the modulations with respect to the lattice sites and analyze the excitation regimes in the weak-drive limit, and we also present a detailed numerical study on the nonlinear behaviors.

Since the lattice system has a finite bandwidth, the particles are expected to eject from the trap when the driving conditions meet the energy conservation, and the emission rate can be reasonably enhanced in the absence of the nonlinearity. The presence of a Peierls phase typically induces imbalanced complex hopping amplitudes, such that varying amounts of excited particles escape from the trap in each direction and counterpropagate along the lattice chain. After passing through the intermediate sites, they continue to travel towards the central site and then re-emit after the collisions with the source condensate. When the number of lattice sites is relatively small, for a site-dependent modulation amplitude there are distinct decaying behaviors between odd and even lattice sites, while the effect does not appear for the case of a uniform amplitude, and the deviations can be greatly reduced with the increase of the lattice sites. The net particle currents along the chain are typically generated by the Peierls phase, and for finite lattice sites they oscillate rapidly around certain values, even when the driving field is turned off after half of the period.

The ring-shaped lattice model is experimentally accessible and is of importance in the physical realization for a generic quantum system of interacting particles. One could routinely boost the particle emission by manipulating the time-periodic driving field, and the maintenance and control of the net particle currents can be achieved by employing typical Peierls phases. Though we focus on the bosonic case, one could also load a mixture of multi-component or multi-species atoms into the lattice, and further vary the modulation amplitude in different ways with quasi-periodic driving fields, which might greatly change the dynamics. In addition, the purely imaginary hopping amplitude and the introduction of next-nearest-neighbor hopping may give rise to rich and interesting nonlinear effects of the tunneling of particle currents between weakly coupled lattices, which are worth investigating, and we leave them for future research.

\section*{Acknowledgements}
We thank Yuyi Xue for fruitful discussions. This work was supported by the Natural Science Research Start-up Foundation of Recruiting Talents of Nanjing University of Posts and Telecommunications (Grant No.~NY223065).

\begin{appendix}
\begin{widetext}
\section{Frequency-domain Green's function} \label{gfunction}
Here we present the detailed derivations of Eq.~(\ref{epsilon}) in the main text, by starting from the mean-field equation of motion for $j\geq 1$,
\begin{eqnarray}
i\dot{b}_{j}(t)=-Je^{-i\phi}b_{j-1}(t)-Je^{i\phi}b_{j+1}(t)+M_{j} b_{j}(t),
\end{eqnarray}
where the corresponding matrix form reads
\begin{eqnarray}
i\frac{d}{dt}
\left(
\begin{array}{cc}
   b_{1}(t)  \\
   b_{2}(t)  \\
   b_{3}(t)  \\
   \vdots    \\
   b_{L}(t)
\end{array}
\right) &=&
-Je^{-i\phi}
\left(
\begin{array}{cc}
    b_{0}(t)   \\
    b_{1}(t)   \\
    b_{2}(t)   \\
    \vdots   \\
    b_{L-1}(t)
\end{array}
\right)
-Je^{i\phi}
\left(
\begin{array}{cc}
    b_{2}(t)  \\
    b_{3}(t)  \\
    b_{4}(t)  \\
    \vdots  \\
    b_{L+1}(t)
\end{array}
\right) \nonumber\\
&&+
\left(
\begin{array}{cc}
    M_{1}b_{1}(t)   \\
    M_{2}b_{2}(t)   \\
    M_{3}b_{3}(t)   \\
    \vdots   \\
    M_{L}b_{L}(t)
\end{array}
\right).
\end{eqnarray}
Applying the Fourier transform $b_{j}(\omega) = \int e^{-i\omega t}b_{j}(t)dt$ and shuffling terms around lead to

\begin{eqnarray}
\left(
\begin{array}{cc}
   b_{1}(\omega)  \\
   b_{2}(\omega)  \\
   b_{3}(\omega)  \\
   \vdots     \\
   b_{L}(\omega)
\end{array}
\right) &=& \left(
\begin{array}{cccccc}
\omega-M_{1} & Je^{i\phi}  & 0 & 0  &  \ldots  \\
Je^{-i\phi} & \omega-M_{2} & Je^{i\phi} & 0  &\ldots      \\
0 & Je^{-i\phi} & \omega-M_{3} & Je^{i\phi}  & \ldots \\
\vdots & \vdots & \vdots & \ddots &  \vdots \\
0 & 0 & 0 & Je^{-i\phi} & \omega-M_{L}
\end{array}
\right)^{-1}
\left(
\begin{array}{cc}
    -Je^{-i\phi}b_{0}(\omega)  \\
     0   \\
     \vdots \\
     0   \\
     -Je^{i\phi}b_{0}(\omega)
\end{array}
\right) \nonumber \\
&&=\left(
\begin{array}{ccccc}
   {\cal G}_{11}  & {\cal G}_{12}  & {\cal G}_{13} & \ldots & {\cal G}_{1L} \\
   {\cal G}_{21}  & {\cal G}_{22}  & {\cal G}_{23} & \ldots &  {\cal G}_{2L}\\
   {\cal G}_{31}  & {\cal G}_{32}  & {\cal G}_{33} & \ldots & {\cal G}_{3L}\\
   \vdots & \vdots & \vdots   & \ddots & \vdots \\
   {\cal G}_{L1} & {\cal G}_{L2} & {\cal G}_{L3} & \ldots & {\cal G}_{LL}
\end{array}
\right)
\left(
\begin{array}{cc}
    -Je^{-i\phi}b_{0}(\omega)  \\
     0   \\
     \vdots \\
     0   \\
     -Je^{i\phi}b_{0}(\omega)
\end{array}
\right) \nonumber \\
&&=\left(
\begin{array}{cc}
      -Je^{-i\phi}b_{0}(\omega){\cal G}_{11}-Je^{i\phi}b_{0}(\omega){\cal G}_{1L}  \\
     -Je^{-i\phi}b_{0}(\omega){\cal G}_{21}-Je^{i\phi}b_{0}(\omega){\cal G}_{2L}   \\
     -Je^{-i\phi}b_{0}(\omega){\cal G}_{31}-Je^{i\phi}b_{0}(\omega){\cal G}_{3L}   \\
     \vdots \\
     -Je^{-i\phi}b_{0}(\omega){\cal G}_{L1}-Je^{i\phi}b_{0}(\omega){\cal G}_{LL}
\end{array}
\right),
\end{eqnarray}
\end{widetext}
with ${\cal G}_{mn}$ being the element of the inversion of the matrix, and a general form of $b_{j}(\omega)$ is thus straightforward
\begin{eqnarray} \label{gform}
b_{j}(\omega)=-Je^{-i\phi}b_{0}(\omega){\cal G}_{j1}-Je^{i\phi}b_{0}(\omega){\cal G}_{jL}.
\end{eqnarray}
According to the fact that
\begin{eqnarray} \label{sequence}
(\omega-M_{j})b_{j}(\omega)=-Je^{-i\phi}b_{j-1}(\omega)-Je^{i\phi}b_{j+1}(\omega),
\end{eqnarray}
collecting the terms proportional to $e^{\pm i\phi}$ yields
\begin{eqnarray}
b_{j-1}(\omega) &=& (\omega-M_{j})b_{0}(\omega){\cal G}_{j1}, \label{Gj1} \\
b_{j+1}(\omega) &=& (\omega-M_{j})b_{0}(\omega){\cal G}_{jL}.
\end{eqnarray}

\subsection{The case of $M_{j}\equiv M$}

We first consider the case of a uniform modulation amplitude $M_{j}\equiv M$, and we directly have
\begin{eqnarray}
{\cal G}_{11}={\cal G}_{LL}=\frac{1}{\omega-M}.
\end{eqnarray}
A nice simplification is to assume $b_{j}(\omega)=\lambda e^{-\kappa(\phi,\omega)j}$; thus for $j=1$ we get
\begin{eqnarray}
(\omega-M)b_{1}(\omega) &=& -Je^{-i\phi}b_{0}(\omega) -Je^{i\phi}b_{2}(\omega),
\end{eqnarray}
and
\begin{eqnarray}
\lambda &=& \frac{-Je^{-i\phi}e^{\kappa(\phi,\omega)}b_{0}(\omega)}{(\omega-M)+Je^{i\phi}e^{-\kappa(\phi,\omega)}}.
\end{eqnarray}
As for site $j>1$,
\begin{eqnarray}
&&(\omega-M)\lambda e^{-\kappa(\phi,\omega)j} \nonumber \\
&=& -Je^{-i\phi}\lambda e^{-\kappa(\phi,\omega)(j-1)}-Je^{i\phi}\lambda e^{-\kappa(\phi,\omega)(j+1)},
\end{eqnarray}
which gives $\lambda=b_{0}(\omega)$, as well as
\begin{eqnarray}
(\omega-M)=-Je^{-i\phi}e^{\kappa}-Je^{i\phi}e^{-\kappa},
\end{eqnarray}
and hence we explicitly obtain
\begin{eqnarray}
\kappa(\phi,\omega)={\rm Arc}\left[\cosh{\left(\frac{\omega-M}{-2J}\right)}\right]+i\phi.
\end{eqnarray}
Now we are in position to analytically calculate ${\cal G}_{1L}$ and ${\cal G}_{L1}$. The direct substitution of the above ansatz of $b_{j}(\omega)$ into Eq.~(\ref{gform}) leads to
\begin{eqnarray}
e^{-\kappa j}=-Je^{-i\phi}{\cal G}_{j1}-Je^{i\phi}{\cal G}_{jL},
\end{eqnarray}
where the neat combination
\begin{eqnarray}
(\omega-M)=J^{2}({\cal G}_{11}+e^{2i\phi}{\cal G}_{1L})+\frac{1}{{\cal G}_{11}+e^{2i\phi}{\cal G}_{1L}}
\end{eqnarray}
gives the analytical expression
\begin{eqnarray}
{\cal G}_{1L}=\frac{(\omega-M)^{2}-2J^{2}\pm\sqrt{(\omega-M)^{4}-4J^{2}(\omega-M)^{2}}}{2J^{2}(\omega-M)e^{2i\phi}}. \nonumber \\
\end{eqnarray}
With respect to ${\cal G}_{L1}$, we start from
\begin{eqnarray}
e^{-\kappa L}=-Je^{-i\phi}{\cal G}_{L1}-Je^{i\phi}{\cal G}_{LL}.
\end{eqnarray}
Taking $j=L$ in Eq.~(\ref{Gj1}), we get
\begin{eqnarray}
e^{-\kappa L}e^{\kappa}=(\omega-M){\cal G}_{L1}
\end{eqnarray}
and reach ${\cal G}_{L1}$ by combination of the above equations:
\begin{eqnarray}
{\cal G}_{L1}=\frac{2J^{2}e^{2i\phi}(\omega-M)^{-1}}{(\omega-M)^{2}-2J^{2}\pm \sqrt{(\omega-M)^{4}-4J^{2}(\omega-M)^{2}}}. \nonumber \\
\end{eqnarray}

\subsection{The case of $M_{j}=(-1)^j$}

Similarly, for the site-dependent case of $M_{j}=(-1)^{j}$, with $j=1$ and $j=L$ we have
\begin{eqnarray}
{\cal G}_{11}=\frac{1}{\omega-M_{1}}, \text{ \ } {\cal G}_{LL}=\frac{1}{\omega-M_{L}}.
\end{eqnarray}
By assuming that
\begin{eqnarray}
b_{j}(\omega)=
\left\{
\begin{array}{cc}
  b_{0}(\omega)e^{-\kappa_{\rm{o}}(\phi,\omega)j},   &  \text{$j$ is odd,}\\
  b_{0}(\omega)e^{-\kappa_{\rm{e}}(\phi,\omega)j},   &  \text{$j$ is even,}
\end{array}
\right.
\end{eqnarray}
and according to Eq.~(\ref{sequence}), we obtain for odd $j$ and even $j$, respectively,
\begin{eqnarray}
(\omega-M_{2n-1})e^{-\kappa_{\rm{o}}j} &=& -Je^{-i\phi}e^{-\kappa_{\rm{e}}(j-1)}-Je^{i\phi}e^{-\kappa_{\rm{e}}(j+1)}, \nonumber \\
(\omega-M_{2n})e^{-\kappa_{\rm{e}}j} &=& -Je^{-i\phi}e^{-\kappa_{\rm{o}}(j-1)}-Je^{i\phi}e^{-\kappa_{\rm{o}}(j+1)}. \nonumber \\
\end{eqnarray}
One can simply consider $j=1$ and $j=2$ for the above equations to find
\begin{eqnarray} \label{kappae}
e^{-2\kappa_{\rm{e}}}=\frac{(\omega-M_{2n-1})e^{-\kappa_{\rm{o}}}+Je^{-i\phi}}{-Je^{i\phi}}
\end{eqnarray}
and
\begin{eqnarray}
J^{2}e^{2i\phi}e^{-3\kappa_{\rm{o}}}&+&\left[J^{2}-(\omega-M_{2n-1})(\omega-M_{2n})\right]e^{-\kappa_{\rm{o}}}\nonumber\\
&&-(\omega-M_{2n})Je^{-i\phi}=0.
\end{eqnarray}
Based on the relations in Eq.~(\ref{gform}), taking $j=1$ we get
\begin{eqnarray}
{\cal G}_{1L}= \frac{e^{-\kappa_{\rm{o}}}+Je^{-i\phi}{\cal G}_{11}}{-Je^{i\phi}}.
\end{eqnarray}
It is apparent that for odd $L$,
\begin{eqnarray}
{\cal G}_{L1}^{\rm{o}}=\frac{e^{-\kappa_{\rm{o}}L}+Je^{i\phi}{\cal G}_{LL}^{\rm{o}}}{-Je^{-i\phi}},
\end{eqnarray}
while for even $L$,
\begin{eqnarray} \label{GL1e}
{\cal G}_{L1}^{\rm{e}}=\frac{e^{-\kappa_{\rm{e}}}L+Je^{i\phi}{\cal G}_{LL}^{\rm{e}}}{-Je^{-i\phi}}.
\end{eqnarray}
We also have the equations of motion for $b_{0}(t)$:
\begin{eqnarray}
i\dot{b}_{0}(t) &=& V_{\rm{tr}}b_{0}(t)+[U+g(t)]|b_{0}(t)|^{2}b_{0}(t) \nonumber\\
&&-Je^{i\phi}b_{1}(t)-Je^{-i\phi}b_{L}(t).
\end{eqnarray}
Substituting the ansatz $b_{0}(t)=\beta e^{-i \epsilon^{\prime} t}$ at $g=0$ and performing the Fourier transform, we reach
\begin{eqnarray} \label{epsilon1}
\epsilon^{\prime}=V_{\rm{tr}}+U|\beta|^{2}-Je^{i\phi}{\cal B}_{1}(\epsilon^{\prime})-Je^{-i\phi}{\cal B}_{L}(\epsilon^{\prime}),
\end{eqnarray}
where
\begin{eqnarray}
{\cal B}_{1}(\epsilon^{\prime}) &=& -Je^{-i\phi}{\cal G}_{11}(\epsilon^{\prime})-Je^{i\phi}{\cal G}_{1L}(\epsilon^{\prime}), \\
{\cal B}_{L}(\epsilon^{\prime}) &=& -Je^{-i\phi}{\cal G}_{L1}(\epsilon^{\prime})-Je^{i\phi}{\cal G}_{LL}(\epsilon^{\prime}). \label{BL}
\end{eqnarray}
Finally, $\epsilon^{\prime}$, $\kappa_{\rm{o}}$ and $\kappa_{\rm{e}}$ can be found by numerically solving Eqs.~(\ref{kappae})-(\ref{GL1e}) and (\ref{epsilon1})-(\ref{BL}) based upon the corresponding odd $L$ and even $L$.

\end{appendix}

\end{document}